\documentclass{article}
\usepackage[T1]{fontenc}
\usepackage{url}
\usepackage[dvipdfmx]{graphicx}
\usepackage{latexsym}
\usepackage[numbers]{natbib}
\usepackage{rotating}
\usepackage{amsmath,amssymb,amsfonts}
\usepackage{algorithm,algorithmic}
\usepackage{textcomp}
\usepackage{booktabs}
\usepackage{multirow}
\usepackage{xcolor}

\begin{document}
%
\title{Acceleration of multicomponent multiple-precision arithmetic with branch-free algorithms and SIMD vectorization}

%
%
\author{Tomonori Kouya\\Otemon Gakuin University\\OrcidID:0000-0003-0178-5519}
%
%
%
\maketitle 
\begin{abstract}
Multiple-precision floating-point branch-free algorithms can significantly accelerate multicomponent arithmetic implemented by combining hardware-based binary64 and binary32, particularly for triple- and quadruple-precision computations. In this study, we achieved benchmark results on x86 and ARM CPU platforms to quantify the accelerations achieved in linear computations and polynomial evaluations by integrating these algorithms.

\end{abstract}
%
\section{Introduction}

Recent advances in artificial intelligence (AI), underpinned by deep-learning technologies, have led to the widespread adoption of large language models and generative AI systems, permeating diverse industrial sectors and everyday human activities, including software development and coding. This trend has accelerated the shortening of the mantissa lengths in floating-point formats employed in AI computations to achieve higher performance. However, in scientific and engineering computing, double-precision arithmetic is the minimum standard, and ill-conditioned problems frequently require higher precision. In such cases, the effective use of single-instruction multiple-data (SIMD) instructions and OpenMP-based parallelization techniques is crucial for harnessing the full capabilities of modern hardware. Among the existing multiple-precision arithmetic frameworks, arbitrary-precision arithmetic is dominated by MPFR~\cite{mpfr}, which relies on the multiple-precision natural-number arithmetic (MPN) kernel of GMP~\cite{gmp}. MPC~\cite{mpc}, an arbitrary precision complex arithmetic library built on MPFR that achieves acceleration through the Karatsuba 3M multiplication algorithm, is widely used.

In contrast, fixed-precision multicomponent arithmetic based on error-free transformation (EFT) techniques, which employs combinations of multiple IEEE binary32 (24-bit mantissa and 32-bit total length) and binary64 values supported by existing hardware, is represented by quadruple-double (QD)~\cite{qd} for CPUs and GQD~\cite{GQD}---a CUDA port of QD, though dated---for GPUs. When a relatively short mantissa length suffices, such as in double-double (DD, 106-bit), triple-double (TD, 159-bit), and quadruple-double (QD, 212-bit) arithmetic implemented by combining two to four binary64 values, computations can be executed faster than with MPFR. Kotakemori et al.'s Lis~\cite{lis} and Hishinuma et al.~\cite{dd_avx_original} further demonstrated that basic linear computations can be accelerated by leveraging AVX, AVX2, and AVX-512, which are x86-based SIMD instruction sets. Building on an existing AVX2 implementation, we demonstrated that basic linear computations in DD, TD, and QD precision can be accelerated on the Raspberry Pi and Snapdragon platforms by exploiting the Neon instructions~\cite{kouya_arm_neon_2025}.

Nevertheless, DD, TD, and QD arithmetic require renormalization at the conclusion of each operation. In particular, for TD and QD arithmetic, this step has been identified as a bottleneck that impedes performance gains through SIMD vectorization. Double-precision pair arithmetic, which bypasses renormalization, has been proposed as a faithfully rounded pair-arithmetic scheme~\cite{pair_arithmetic}, and practical sparse matrix solver implementations that exploit this efficiency have been reported~\cite{pair_arith_sparse2025}. The branch-free algorithms for multiple-precision floating-point arithmetic addressed in this work~\cite{branch_free_2025} eliminate all conditional branches from every algorithmic step, including renormalization, and reduce the operation count for TD and QD arithmetic relative to conventional algorithms. Consequently, this methodology can substantially accelerate triple- and quadruple-precision multicomponent arithmetic implemented by combining hardware-based binary64 and binary32 values and is immediately applicable to existing fixed-precision multiple-precision arithmetic implementations. Because the benchmarks presented in the original paper provided an insufficient comparison against existing SIMD implementations, we first quantified the speedups achievable in basic linear computations by incorporating these algorithms on the x86 and ARM CPU platforms. Furthermore, as a practical application, we analyzed the effectiveness of the proposed approach by applying it to a multiple-precision algebraic equation solver.

%
\section{Branch-Free Algorithms for DD, TD, and QD Addition and Multiplication}

The extension of fixed-precision floating-point numbers via EFT was pioneered by Dekker~\cite{dekker}. Multiple-precision arithmetic with an extended mantissa length was implemented by employing EFT primitives, represented by QuickTwoSum (Algorithms~\ref{algo:QuickTwoSum}, \ref{algo:TwoSum}, \ref{algo:TwoProd-fma}), which return both a rounded result and its rounding error using only existing floating-point operations (addition~$\oplus$, subtraction~$\ominus$, and multiplication~$\otimes$). In the following, all variables are assumed to be exactly representative as normalized floating-point numbers.

QuickTwoSum (Algorithm~\ref{algo:QuickTwoSum}) returns the floating-point addition result $s = a \oplus b$ and its rounding error $e$, but only under the precondition $|a| \ge |b|$; if no overflow occurs, it maintains the relation $a + b = s + e$.

\begin{algorithm}[htb]
\caption{($s$, $e$) $:=$ QuickTwoSum($a$, $b$)}\label{algo:QuickTwoSum}
\begin{algorithmic}
	\STATE $s :=a \oplus b$; $e := b \ominus (s \ominus a)$; \textbf{return} ($s$, $e$)
\end{algorithmic}
\end{algorithm}

TwoSum (Algorithm~\ref{algo:TwoSum}) returns the floating-point addition result $s = a \oplus b$ and its rounding error $e$ unconditionally, regardless of the relative magnitudes of $a$ and $b$, while maintaining the relationship $a + b = s + e$.
\begin{algorithm}[htb]
\caption{($s$, $e$) $:=$ TwoSum($a$, $b$)}\label{algo:TwoSum}
\begin{algorithmic}
	\STATE $s := a \oplus b$; $v := s \ominus a$; $e := (a \ominus (s \ominus v)) \oplus (b \ominus v)$; \textbf{return}($s$, $e$)
\end{algorithmic}
\end{algorithm}

TwoProd (Algorithm~\ref{algo:TwoProd-fma}) returns the floating-point multiplication result $p = a \otimes b$ and its rounding error $e$ while maintaining the relationship $ab = p + e$. Because virtually all floating-point hardware currently in operation supports fused multiply–add (FMA) instructions, the FMA-based variant of TwoProd was employed throughout this work.

\begin{algorithm}[htb]
\caption{($p$, $e$) $:=$ TwoProd($a$, $b$)}\label{algo:TwoProd-fma}
\begin{algorithmic}
	\STATE $p := a \otimes b$; $e := \mbox{FMA}(a, b, -p)$ ($= a \times b - p$); \textbf{return}($p$, $e$)
\end{algorithmic}
\end{algorithm}

As these EFT operations are constructed entirely from existing floating-point operations and contain no conditional branches, they can be implemented directly using SIMD instructions. With AVX2, which employs 256-bit registers on x86 architectures, four binary64 or eight binary32 values can be processed simultaneously by EFT operations. In contrast, with Neon on AArch64, which employs 128-bit registers, two binary64 or four binary32 values can be processed simultaneously. Computations that contain no conditional branches and can be executed without overflowing SIMD registers in this manner are termed branch-free (BF).

By combining EFT operations, multiple-precision arithmetic with extended mantissa can be implemented by composing multiple floating-point numbers; this is referred to as the multicomponent (or multiterm) method. Because subtraction reduces to sign negation and division can be implemented via Newton's method that combines addition and multiplication, these two operations are crucial from a performance perspective. Correct handling of the rounding errors is critical in this implementation and requires a carefully derived rigorous error analysis. As the mantissa length increases, the operation count increases, and error management becomes increasingly complex. Zhang \& Aiken proposed an abstract model for automated reasoning that simplifies floating-point arithmetic~\cite{automatic_verification_2025} and introduced branch-free multicomponent multiple-precision addition and multiplication algorithms. In the following section, we present and discuss BF algorithms.

\subsection{Remarks on double-word (DW) Arithmetic}

The simplest form of multicomponent multiple-precision floating-point arithmetic is the DW arithmetic. When binary64 (double-precision) is used as the base format, it is abbreviated as DD; when binary32 (single-precision) is used, it is abbreviated as DS. Algorithm~\ref{algo:ddadd} presents the default DW addition defined in QD~\cite{qd}, which is the sloppy or fast variant; its final rounding error is larger than that of the accurate version (Algorithm~\ref{algo:ddaccurateadd}).

\begin{algorithm}[htb]
\caption{$c[2] :=$ DWAdd($a[2]$, $b[2]$)}\label{algo:ddadd}
\begin{algorithmic}
	\STATE $(s, e) := \mbox{TwoSum}(a[0], b[0])$; $e := e \oplus (a[1] \oplus b[1])$; $(c[0], c[1]) := \mbox{QuickTwoSum}(s, e)$;
	\STATE \textbf{return}($c[2]$)
\end{algorithmic}
\end{algorithm}

\begin{algorithm}[htb]
\caption{$c[2] :=$ DWAccurateAdd($a[2]$, $b[2]$)}\label{algo:ddaccurateadd}
\begin{algorithmic}
	\STATE $(s_1, s_2) := \mbox{TwoSum}(a[0], b[0])$; $(t_1, t_2) := \mbox{TwoSum}(a[1], b[1])$;
	\STATE $s_2 := s_2 \oplus t_1$; $(s_1, s_2) := \mbox{QuickTwoSum}(s1, s2)$;
	\STATE $s_2 := s_2 \oplus t_2$; $(c[0], c[1]) := \mbox{QuickTwoSum}(s1, s2)$;
	\STATE \textbf{return}($c[2]$)
\end{algorithmic}
\end{algorithm}

Zhang and Aiken proposed Algorithm~\ref{algo:ddaddbf} as a branch-free reformulation for accurate DW addition.
	
\begin{algorithm}[htb]
\caption{$c[2] :=$ DWBFAdd($a[2]$, $b[2]$)}\label{algo:ddaddbf}
\begin{algorithmic}
	\STATE $(g_1, g_{1e}) := \mbox{TwoSum}(a[0], b[0])$; \STATE $(g_2, g_{2e}) := \mbox{TwoSum}(a[1], b[1])$;
	\STATE $(g_3, g_{3e}) := \mbox{QuickTwoSum}(g_1, g_2)$; \STATE $g_4 := g_{1e} \oplus g_{2e}$;
	\STATE $g_5 := g_{4} \oplus g_{3e}$; \STATE $(c[0], c[1]) := \mbox{QuickTwoSum}(g_3, g_5)$;
	\STATE \textbf{return}($c[2]$)
\end{algorithmic}
\end{algorithm}

Algorithm~\ref{algo:ddmul} presents the DW multiplication defined as the default in QD.
\begin{algorithm}[htb]
\caption{$c[2] :=$ DWMul($a[2]$, $b[2]$)}\label{algo:ddmul}
\begin{algorithmic}
	\STATE $(p_1, p_2) := \mbox{TwoProd}(a[0], b[0])$; $p_2 := p_2 \oplus (a[0] \otimes b[1] \oplus a[1] \otimes b[0])$;
	\STATE $(c[0], c[1]) := \mbox{QuickTwoProd}(p_1, p_2)$;
	\STATE \textbf{return}($c[2]$)
\end{algorithmic}
\end{algorithm}

The corresponding BF multiplication is presented as Algorithm~\ref{algo:ddmulbf}. This algorithm can be interpreted as performing all necessary multiplications, followed by arranging the error-handling steps in a branch-free manner.

\begin{algorithm}[htb]
\caption{$c[2] :=$ DWBFMul($a[2]$, $b[2]$)}\label{algo:ddmulbf}
\begin{algorithmic}
 \STATE $(p_{00}, p_{e00}) := \mbox{TwoProd}(a[0], b[0])$; \STATE $p_{01} := a[0] \otimes b[1]$;
 \STATE $ p_{10} = a[1] \otimes b[0]$; \STATE $g_1 := p_{01} \oplus p_{10}$;
 \STATE $g_2 := p_{e00} \oplus g_1$; \STATE $(c[0], c[1]) := \mbox{QuickTwoSum}(p_{00}, g_2)$;
 \STATE \textbf{return}($c[2]$)
\end{algorithmic}
\end{algorithm}

Consequently, the BF variants of DW addition and multiplication involve no fewer floating-point operations than their conventional counterparts. In addition, as discussed in the subsequent sections, they exhibit no measurable performance benefit under the current default DW arithmetic. Therefore, in the following benchmark tests, DD arithmetic employs conventional (non-BF) implementations unless otherwise stated; only TD and QD arithmetic utilize the BF algorithms presented hereafter.

\subsection{Remarks on TW Arithmetic}

For triple-precision arithmetic (triple-word, TW), one approach is to consider the fourth component of the QD arithmetic as zero. In contrast, Fabiano et al.~\cite{triple_word_prec2019} proposed a formulation with a reduced operation count. Through SIMD implementation, we validated that the former approach outperforms the latter~\cite{kouya_iccsa2021}; therefore, Algorithm~\ref{algo:tdadd} is used for TW addition and Algorithm~\ref{algo:tdmul} for TW multiplication.

\begin{algorithm}[htb]
\caption{$c[3] :=$ TWAdd($a[3]$, $b[3]$)}\label{algo:tdadd}
\begin{algorithmic}
	\STATE $s_0 := a[0] \oplus b[0]$; $s_1 := a[1] \oplus b[1]$; $s_2 := a[2] \oplus b[2]$;
	\STATE $v_0 := s_0 \ominus a[0]$;$v_1 := s_1 \ominus a[1]$;$v_2 := s_2 \ominus a[2]$;
	\STATE $u_0 := s_0 \ominus v_0$; $u_1 := s_1 \ominus v_1$; $u_2 := s_2 \ominus v_2$;
	\STATE $w_0 := a[0] \ominus u_0$;$w_1 := a[1] \ominus u_1$; $w_2 := a[2] \ominus u_2$;
	\STATE $u_0 := b[0] \ominus v_0$; $u_1 := b[1] \ominus v_1$; $u_2 := b[2] \ominus v_2$;
	\STATE $t_0 := w_0 \oplus u_0$; $t_1 := w_1 \oplus u_1$; $t_2 := w_2 \oplus u_2$;
	\STATE $(s_1, t_0) := \mbox{TwoSum}(s_1, t_0)$; $(i_1, i_2) := \mbox{TwoSum}(s_2, t_0)$;
	\STATE $(s_2, i_3) := \mbox{TwoSum}(t_1, i_1)$; $(t_0, t_1) := \mbox{TwoSum}(i_2, i_3)$ $t_0 := t_0 \oplus t_1 \oplus t_2$;
	\STATE $c[3] := \mbox{TWRenormalize}(s_0, s_1, s_2, t_0)$;
	\STATE \textbf{return}($c[3]$)
\end{algorithmic}
\end{algorithm}

\begin{algorithm}[htb]
\caption{$c[3] :=$ TWMul($a[3]$, $b[3]$)}\label{algo:tdmul}
\begin{algorithmic}
	\STATE $(p_0, q_0) := \mbox{TwoProd}(a[0], b[0])$; $(p_1, q_1) := \mbox{TwoProd}(a[0], b[1])$;
	\STATE $(p_2, q_2) := \mbox{TwoProd}(a[1], b[0])$;
	\STATE $(p_3, q_3) := \mbox{TwoProd}(a[0], b[2])$; $(p_4, q_4) := \mbox{TwoProd}(a[1], b[1])$;
	\STATE $(p_5, q_5) := \mbox{TwoProd}(a[2], b[0])$;
	\STATE $(i_1, i_2) := \mbox{TwoSum}(p_1, p_2)$; $(p_1, i_3) := \mbox{TwoSum}(q_0, i_1)$; $(p_2, q_0) := \mbox{TwoSum}(i_2, i_3)$;
	\STATE $(i_1, i_2) := \mbox{TwoSum}(p_2, q_1)$; $(p_2, i_3) := \mbox{TwoSum}(q_2, i_1)$; $(q_1, q_2) := \mbox{TwoSum}(i_2, i_3)$;
	\STATE $(i_1, i_2) := \mbox{TwoSum}(p_3, p_4)$; $(p_3, i_3) := \mbox{TwoSum}(p_5, i_1)$; $(p_4, p_5) := \mbox{TwoSum}(i_2, i_3)$;	
	\STATE $(s_0, t_0) := \mbox{TwoSum}(p_2, p_3)$; $(s_1, t_1) := \mbox{TwoSum}(q_1, p_4)$; $s_2 := q_2 \oplus p_5$;
	\STATE $(s_1, t_0) := \mbox{TwoSum}(s_1, t_0)$; $s_2 := s_2 \oplus (t_0 \oplus t_1)$; $(q_0, q_3) := \mbox{TwoSum}(q_0, q_3)$;
	\STATE $(q_4, q_5) := \mbox{TwoSum}(q_4, q_5)$; $(t_0, t_1) := \mbox{TwoSum}(q_0, q_4)$; $t_1 := t_1 \oplus (q_3 \oplus q_5)$;
	\STATE $(t_0, t_1) := \mbox{TwoSum}(q_3, s_1)$;
	\STATE $c[3] := \mbox{TWRenormalize}(p_0, p_1, s_0, t_0)$;
	\STATE \textbf{return}($c[3]$)
\end{algorithmic}
\end{algorithm}

Both algorithms conclude with a call to the TW renormalization function TWRenormalize\cite{triple_word_prec2019}\cite{qd}, which requires up to five QuickTwoSum operations and necessitates conditional branching via if-statements. This function has not yet been SIMD-vectorized and is considered the primary source of performance degradation.

By contrast, the proposed BF algorithms presented in Algorithms~\ref{algo:tdaddbf} and~\ref{algo:tdmulbf} introduce no conditional branches in the renormalization-equivalent steps and require fewer operations than their conventional counterparts. This renders them highly amenable to performance improvements through SIMD vectorization.

\begin{algorithm}[htb]
\caption{$c[3] :=$ TWBFAdd($a[3]$, $b[3]$)}\label{algo:tdaddbf}
\begin{algorithmic}
 \STATE $(a_1, b_1) := \mbox{TwoSum}(a[0], b[0])$; $(c_1, d_1) := \mbox{TwoSum}(a[1], b[1])$;
 \STATE $(e_1, f_1) := \mbox{TwoSum}(a[2], b[2])$;
 \STATE $(a_2, c_2) := \mbox{QuickTwoSum}(a_1, c_1)$; $b_2 := b_1 \oplus f_1$;
 \STATE $(d_2, e_2) := \mbox{TwoSum}(d_1, e_1)$; $(a_3, d_3) := \mbox{QuickTwoSum}(a_2, d_2)$;
 \STATE $(b_3, c_3) := \mbox{TwoSum}(b_2, c_2)$; $c_4 := c_3 \oplus e_2$;
 \STATE $(c_5, d_5) := \mbox{TwoSum}(c_4, d_3)$; $(b_6, c_6) := \mbox{TwoSum}(b_3, c_5)$;
 \STATE $(c[0], b_7) = \mbox{QuickTwoSum}(a_3, b_6)$; $c_7 := c_6 \oplus d_5$;
 \STATE $(c[1], c[2]) := \mbox{QuickTwoSum}(b_7, c_7)$;
	\STATE \textbf{return}($c[3]$)
\end{algorithmic}
\end{algorithm}

\begin{algorithm}[htb]
\caption{$c[3] :=$ TWBFMul($a[3]$, $b[3]$)}\label{algo:tdmulbf}
\begin{algorithmic}
	\STATE $(a_0, b_0) := \mbox{TwoProd}(a[0], b[0])$; $(c_0, e_0) := \mbox{TwoProd}(a[0], b[1])$;
	\STATE $(d_0, f_0) := \mbox{TwoProd}(a[1], b[0])$;
	\STATE $g_0 := a[0] \otimes b[2]$; $h_0 := a[1] \otimes b[1]$; $i_0 := a[2] \otimes b[0]$;
	\STATE $(c_1, d_1) := \mbox{TwoSum}(c_0, d_0)$; $e_1 := e_0 \oplus f_0$; $g_1 := g_0 \oplus i_0$;
	\STATE $(b_2, c_2) := \mbox{TwoSum}(b_0, c_1)$; $g_2 := g_1 \oplus h_0$; $(a_3, b_3) := \mbox{QuickTwoSum}(a_0, b_2)$;
	\STATE $c_3 := c_2 \oplus d_1$; $e_3 := e_1 \oplus g_2$; $c_4 := c_3 \oplus e_3$;
	\STATE $(b_5, c_5) := \mbox{QuickTwoSum}(b_3, c_4)$; $(c[0], b_6) := \mbox{QuickTwoSum}(a_3, b_5)$;
	\STATE $(c[1], c[2]) := \mbox{QuickTwoSum}(b_6, c_5)$;
	\STATE \textbf{return}($c[3]$)
\end{algorithmic}
\end{algorithm}

As listed in \tablename\ \ref{table:twcount}, the branch-free triple-word multiplication has the least total number of primary computations compared with our currently adopted and Fabiano's and triple-word multiplications. However, we could not find a clear difference among the three types of triple-word additions.

\begin{table}[htb]
\centering
\caption{Operation counts for Triple-Word arithmetic}\label{table:twcount}
\begin{tabular}{l|rrrrr}
\toprule
Triple-Word (TW) & Add & Sub & Mul & Div & FMA \\
\midrule
TW+TW & 19 & 34 & 0 & 0 & 0 \\
Fabiano TW+TW & 20 & 40 & 0 & 0 & 0 \\
BF TW+TW & 21 & 36 & 0 & 0 & 0 \\\midrule
TW$\times$TW & 42 & 74 & 6 & 0 & 6 \\
Fast Fabiano TW$\times$TW & 24 & 38 & 0 & 0 & 3 \\
BF TW$\times$TW & 14 & 16 & 6 & 0 & 3 \\
\bottomrule
\end{tabular}
\end{table}

\subsection{Remarks on QW Arithmetic}

Quadruple-precision arithmetic (quadruple-word, QW) is based on the implementation provided in QD. As noted previously, TD arithmetic is an abbreviated form of QD arithmetic with the fourth component zeroed. Both the QD addition and multiplication conclude with a call to the QW renormalization function (QWRenormalize\cite{qd}), which contains conditional branches and requires up to seven QuickTwoSum function calls.

The BF algorithms are presented in Algorithms~\ref{algo:qdaddbf} and~\ref{algo:qdmulbf}. Although the time required to discover these BF algorithms is unknown, the resulting algorithms are remarkably concise and involve substantially reduced operation counts. Consequently, the benefit of SIMD vectorization is expected to be more pronounced for QW arithmetic than for TW arithmetic.

\begin{algorithm}[htb]
\caption{$c[4] :=$ QWBFAdd($a[4]$, $b[4]$)}\label{algo:qdaddbf}
\begin{algorithmic}
 \STATE $(a_1, b_1) := \mbox{TwoSum}(a[0], b[0])$; $(c_1, d_1) := \mbox{TwoSum}(a[1], b[1])$;
 \STATE $(e_1, f_1) := \mbox{TwoSum}(a[2], b[2])$;
 \STATE $(g_1, h_1) := \mbox{TwoSum}(a[3], b[3])$; $(a_2, c_2) := \mbox{QuickTwoSum}(a_1, c_1)$;
 \STATE $b_2 := b_1 \oplus h_1$; $(d_2, e_2) := \mbox{TwoSum}(d_1, e_1)$;
 \STATE $(f_2, g_2) := \mbox{TwoSum}(f_1, g_1)$; $(b_3, g_3) := \mbox{TwoSum}(b_2, g_2)$;
 \STATE $(c_3, d_3) := \mbox{QuickTwoSum}(c_2, d_2)$; $(e_3, f_3) := \mbox{TwoSum}(e_2, f_2)$;
 \STATE $(a_4, c_4) := \mbox{QuickTwoSum}(a_2, c_3)$; $(d_4, e_4) := \mbox{QuickTwoSum}(d_3, e_3)$;
 \STATE $(b_5, d_5) := \mbox{TwoSum}(b_3, d_4)$; $e_5 := e_4 \oplus f_3$;
 \STATE $(b_6, c_6) := \mbox{TwoSum}(b_5, c_4)$; $(d_6, e_6) := \mbox{TwoSum}(d_5, e_5)$;
 \STATE $(a_7, b_7) := \mbox{QuickTwoSum}(a_4, b_6)$; $(c_7, d_7) := \mbox{QuickTwoSum}(c_6, d_6)$;
 \STATE $e_8 := e_6 \oplus g_3$; $(b_8, c_8) := \mbox{QuickTwoSum}(b_7, c_7)$;
 \STATE $d_9 := d_7 \oplus e_8$; $(c[0], b_{10}) := \mbox{QuickTwoSum}(a_7, b_8)$;
 \STATE $(c_{10}, d_{10}) := \mbox{QuickTwoSum}(c_8, d_9)$; $(c[1], c_{11}) := \mbox{QuickTwoSum}(b_{10}, c_{10})$;
 \STATE $(c[2], c[3]) := \mbox{QuickTwoSum}(c_{11}, d_{10})$;
	\STATE \textbf{return}($c[4]$)
\end{algorithmic}
\end{algorithm}

\begin{algorithm}[htb]
\caption{$c[4] :=$ QDBFMul($a[4]$, $b[4]$)}\label{algo:qdmulbf}
\begin{algorithmic}
 \STATE $(a_0, b_0) := \mbox{TwoProd}(a[0], b[0])$; $(c_0, e_0) := \mbox{TwoProd}(a[0], b[1])$;
 \STATE $(d_0, f_0) := \mbox{TwoProd}(a[1], b[0])$; $(g_0, j_0) := \mbox{TwoProd}(a[0], b[2])$;
 \STATE $(h_0, k_0) := \mbox{TwoProd}(a[1], b[1])$ $(i_0, l_0) := \mbox{TwoProd}(a[2], b[0])$;
 \STATE $m_0 := a[0] \otimes b[3]$; $n_0 := a[1] \otimes b[2]$; $o_0 := a[2] \otimes b[1]$;
 \STATE $p_0 := a[3] \otimes b[0]$; $(c_1, d_1) := \mbox{TwoSum}(c_0, d_0)$;
 \STATE $(e_1, f_1) := \mbox{TwoSum}(e_0, f_0)$; $(g_1, i_1) := \mbox{TwoSum}(g_0, i_0)$;
 \STATE $j_1 := j_0 \oplus l_0$; $m_1 := m_0 \oplus p_0$; $n_1 := n_0 \oplus o_0$;
 \STATE $(b_2, c_2) := \mbox{TwoSum}(b_0, c_1)$; $(e_2, h_2) := \mbox{TwoSum}(e_1, h_0)$;
 \STATE $f_2 := f_1 \oplus j_1$; $i_2 := i_1 \oplus k_0$; $m_2 := m_1 \oplus n_1$;
 \STATE $(a_3, b_3) := \mbox{QuickTwoSum}(a_0, b_2)$; $(c_3, d_3) := \mbox{QuickTwoSum}(c_2, d_1)$;
 \STATE $(e_3, g_3) := \mbox{TwoSum}(e_2, g_1)$; $f_3 := f_2 \oplus m_2$; $h_3 := h_2 \oplus i_2$;
 \STATE $(c_4, e_4) := \mbox{TwoSum}(c_3, e_3)$; $d_4 := d_3 \oplus h_3$; $f_4 := f_3 \oplus g_3$;
 \STATE $d_5 := d_4 \oplus e_4$; $(c_6, d_6) := \mbox{TwoSum}(c_4, d_5)$; $(b_7, c_7) := \mbox{TwoSum}(b_3, c_6)$;
 \STATE $d_7 := d_6 \oplus f_4$; $(c[0], b_8) := \mbox{QuickTwoSum}(a_3, b_7)$;
 \STATE $(c_8, d_8) := \mbox{TwoSum}(c_7, d_7)$; $(c[1], c_9) := \mbox{TwoSum}(b_8, c_8)$;
 \STATE $(c[2], c[3]):= \mbox{QuickTwoSum}(c_9, d_8)$;
	\STATE \textbf{return}($c[4]$)
\end{algorithmic}
\end{algorithm}

We present \tablename\ \ref{table:qwcount} for comparison with our adopted and branch-free quadruple-word algorithms. Similarly, as listed in \tablename\ \ref{table:twcount}, branch-free multiplication is the most lightweight, and no significant difference exists between the two types of quadruple-word additions.

\begin{table}[htb]
\centering
\caption{Operation counts for quad-word arithmetic}\label{table:qwcount}
\begin{tabular}{l|rrrrr}
\toprule
Quad-Word (QW) & Add & Sub & Mul & Div & FMA \\
\midrule
QW+QW & 26 & 58 & 0 & 0 & 0 \\
BF QW+QW & 37 & 66 & 0 & 0 & 0 \\\midrule
QW$\times$QW & 67 & 94 & 12 & 0 & 9 \\
BF QW$\times$QW & 38 & 52 & 10 & 0 & 6 \\
\bottomrule
\end{tabular}
\end{table}

%
\section{Benchmark Tests for Real and Complex Square Matrix Multiplication}

As discussed previously, the adoption of BF arithmetic is expected to yield significant performance gains through SIMD vectorization for triple- and quadruple-precision computations. Here, we report performance evaluation results for real and complex square matrix multiplication obtained by integrating BF arithmetic into our multiple-precision basic linear computation library BNCmatmul~\cite{bncmatmul}, executed on the following x86 AVX2 and Arm Neon environments.

\begin{description}
	\item[EPYC]AMD EPYC 9354P 3.7 GHz 32 cores, Ubuntu 20.04.6 LTS, Intel Compiler version 2021.10.0, GNU MP 6.2.1, MPFR 4.1.0, MPC 1.2.1 
	\item[Snapdragon] Lenovo IdeaPad Slim 5 (Snapdragon X Plus X1P-42-100, 3.4 GHz, 34 MB L3 cache, 8 cores), Windows 11, WSL2 + Ubuntu 24.04.2 LTS, GCC 13.3.0
\end{description}

Standard intrinsic functions, provided as C functions, were used for SIMD implementation in both environments. Arm Neon vectorization and OpenMP parallelization were implemented primarily with assistance from the Anthropic Claude by converting the existing AVX2 code~\cite{kouya_arm_neon_2025}.

%
\subsection{Real Matrix Multiplication}

The square matrices employed in the tests are defined as follows:
\begin{equation}
\begin{split}
 C &:= A B, \\
 A &= [a_{ij}]_{i,j=1, 2, ..., n}, B = [b_{ij}]_{i,j=1, 2, ..., n} \in \mathbb{R}^{n\times n} \\
 a_{ij} &:= \sqrt{5} (i + j - 1), b_{ij} := \sqrt{3} (n - i + 1)
\end{split} \label{eqn:testmatmul}
\end{equation}
as defined in (\ref{eqn:testmatmul}).

Our library supports four matrix multiplication schemes: the naive triple-loop method, a block-based algorithm, Strassen's algorithm, and the Ozaki scheme. In this study, Strassen's algorithm was employed to isolate the performance contribution of the BF arithmetic. The recursion in Strassen's algorithm terminates when the submatrix size decreases below $32 \times 32$. The performance benefit of BF arithmetic was measured for both nonvectorized and SIMD-vectorized (AVX2 and Neon) implementations. Execution times with OpenMP parallelization using 8 and 32 threads were also reported. Because our Strassen routine did not achieve further speedup beyond eight threads, the results for the block-based algorithm (B32 threads) were included only for the 32-thread case. The numerical accuracy of all computed results falls within the ranges of 29.0--30.4 significant decimal digits for DD, 45.5--47.5 for TD, and 61.7--63.5 for QD precision.

\begin{figure*}[htb]
\begin{center}
\includegraphics[width=.49\textwidth]{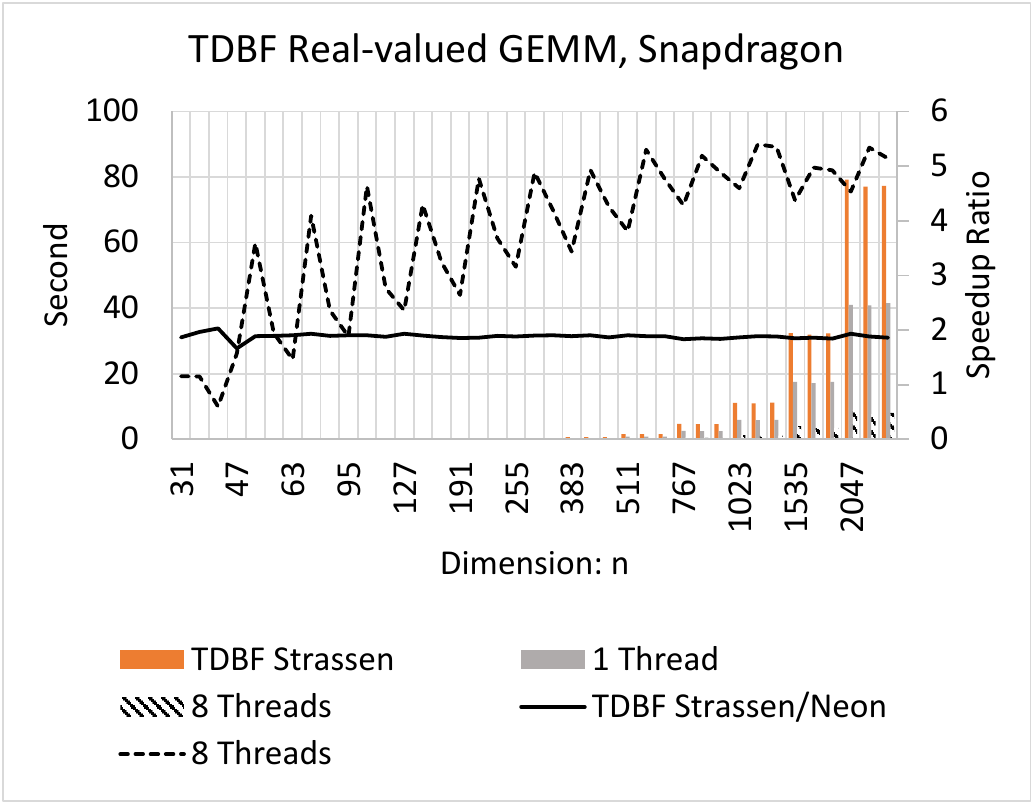}
\includegraphics[width=.49\textwidth]{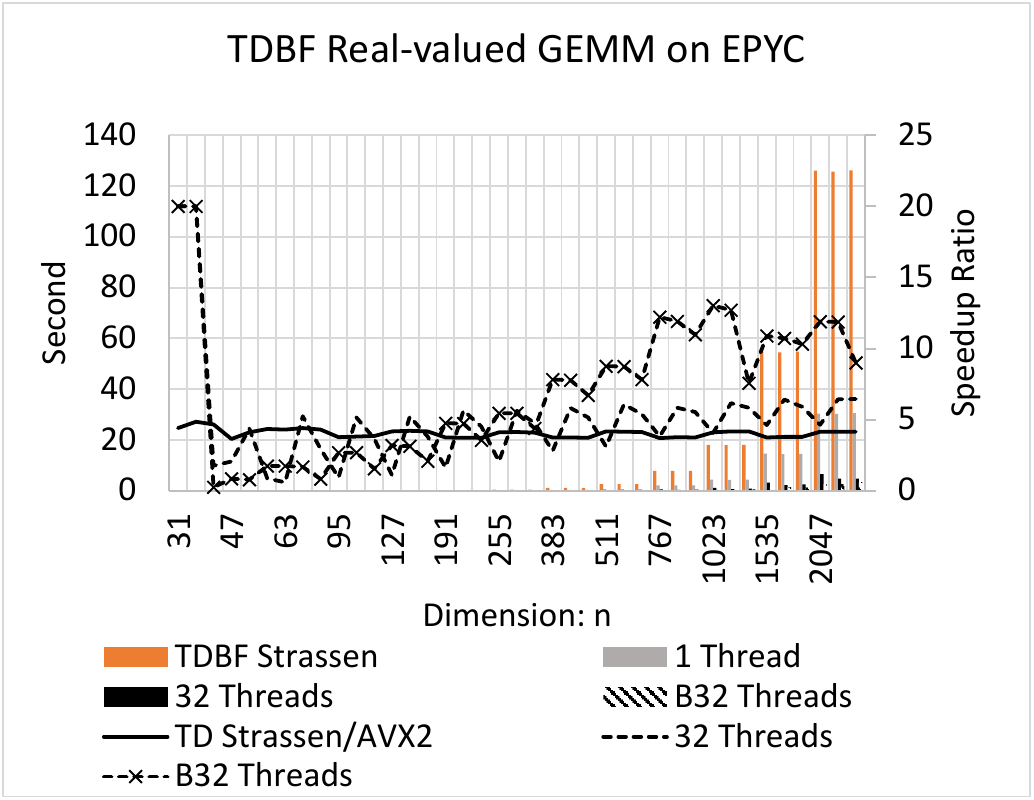}
\includegraphics[width=.49\textwidth]{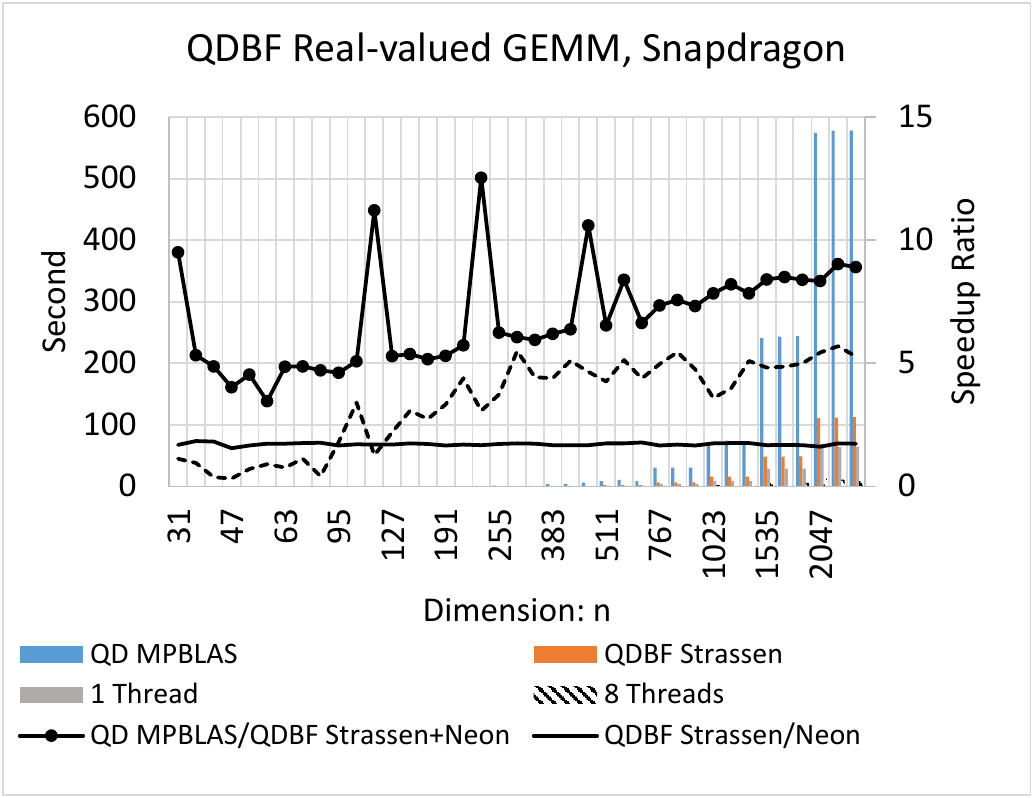}
\includegraphics[width=.49\textwidth]{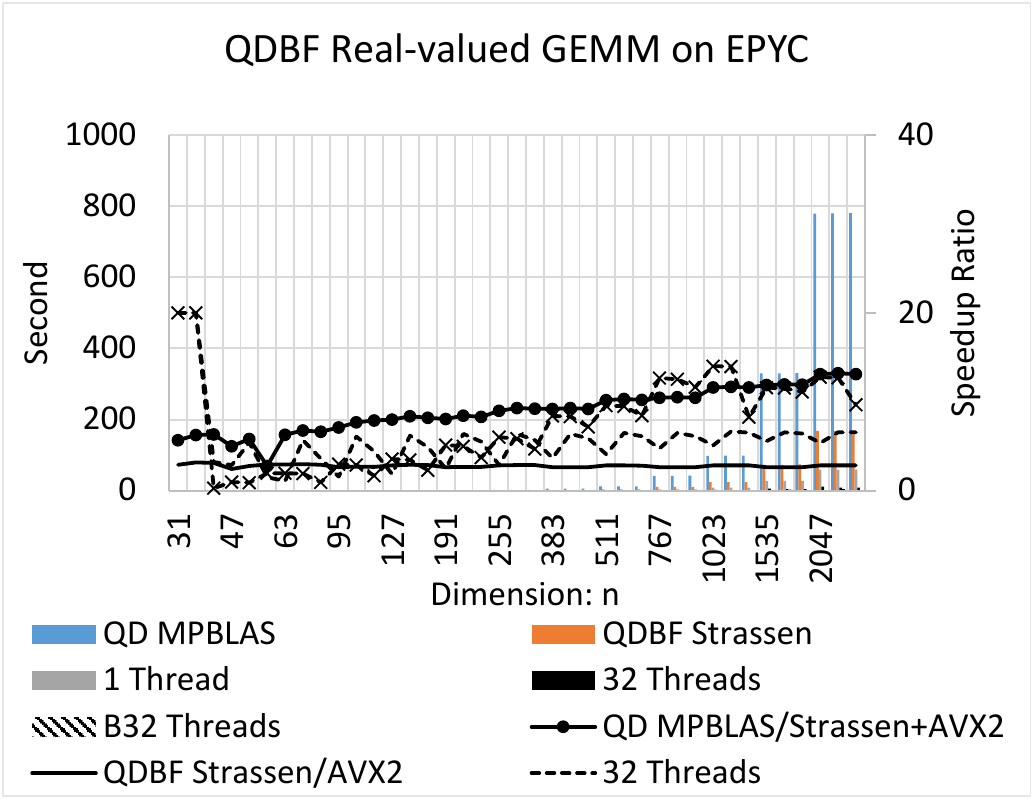}
\caption{Computation time (s) of real-valued Strassen matrix multiplication: Snapdragon (left), EPYC (right)}\label{fig:matmul_strassen_bf}
\end{center}
\end{figure*}

Figure~\ref{fig:matmul_strassen_bf} shows the computation times for real square matrix multiplication at TD and QD precision. The horizontal axis represents the matrix size $n = 32$--$2049$, and the vertical axis represents the computation time in seconds.

Notably, BF arithmetic provides no benefit and degrades performance for DD precision, whereas it yields clear speedups for TD and QD precisions. The resulting speedups over the MPBLAS~\cite{mplapack} DD- and QD-precision Rgemm for $n = 2049$ are summarized in Table~\ref{table:mblas_speedup}.

\begin{table}[htb]
	\begin{center}\small
		\caption{Real-valued matrix multiplication time (s) for $n=2049$}\label{table:mblas_speedup}
\begin{tabular}{ll|r|r|r|r|r|r}
\toprule
 & & MPBLAS & S+SIMD & BF S+SIMD & 8 Threads & 32 Threads & B32 Threads \\
\midrule
\multirow{2}{*}{DD}
& Snapdragon & 13.6 & \underline{9.7} & 15.4 & 2.5 & -- & -- \\
& EPYC & 13.9 & \underline{6.7} & 10.3 & 1.5 & 1.4 & 0.9 \\
\midrule
\multirow{2}{*}{TD}
& Snapdragon & -- & 61.7 & \underline{39.9} & 8.1 & -- & -- \\
& EPYC & -- & 46.1 &\underline{30.4} & 4.8 & 4.7 & 3.4 \\
\midrule
\multirow{2}{*}{QD}
& Snapdragon & 577.5 & 101.4 & \underline{65.6} & 12.3 & -- & -- \\
& EPYC & 781.0 & 117.2 &\underline{59.9} & 9.2 & 9.1 & 6.2 \\
\bottomrule
\end{tabular}
	\end{center}
\end{table}

Table~\ref{table:average_speedup} lists the average speedup ratios owing to BF arithmetic for DD precision, where BF arithmetic offers no benefit compared to nonSIMD (NS) and SIMD-vectorized implementations.

\begin{table}[htb]
	\begin{center}
		\caption{Average speedup with branch-free (BF) arithmetic: standard / BF}\label{table:average_speedup}
		\begin{tabular}{c|cc|cc|cc}
 \toprule
		 Speedup & \multicolumn{2}{c|}{DD} & \multicolumn{2}{c|}{TD} & \multicolumn{2}{c}{QD} \\
 \midrule
			ratio & NS & SIMD & NS & SIMD & NS & SIMD \\ \midrule
			Snapdragon & 0.6 & 0.7 & 1.2 &	1.4 & 2.0 & 1.4 \\ 
			EPYC & 1.0 & 0.7 & 1.4 & 1.5 & 1.5 & 1.9 \\
 \bottomrule
 \end{tabular}
	\end{center}
\end{table}

As expected, DD precision exhibited a slowdown of 0.6--1.0$\times$; however, TD precision achieved speedups of 1.2--1.5$\times$ and QD precision achieved speedups of 1.4--2.0$\times$. Although the true benefit of BF arithmetic is difficult to discern from the benchmarks in the original study, the aforementioned results demonstrate that BF arithmetic effectively accelerates TD and QD precision computations. Furthermore, speedups were achieved through Arm Neon vectorization and OpenMP-based parallelization.

%
\subsection{Benchmark Tests for Complex Matrix Multiplication}

Owing to the previously demonstrated superiority of BF arithmetic for TD and QD computations, complex matrix multiplication benchmarks employed BF-based implementations for comparison. Complex square matrix multiplication is defined using a uniform random variable $r_u \in [0,1]$ and a standard normal random variable $r_n \in \mathbb{R}$ as
\begin{equation}
\begin{split}
 C &:= A B, \\
 A &= [a_{ij}]_{i,j=1, 2, ..., n}, B = [b_{ij}]_{i,j=1, 2, ..., n} \in \mathbb{C}^{n\times n} \\
 a_{ij} &:= \exp(r_n) (r_u - 1/2) + \exp(r_n) (r_u - 1/2) \sqrt{-1} \\
 b_{ij} &:= \exp(r_n) (r_u - 1/2) + \exp(r_n) (r_u - 1/2) \sqrt{-1}
\end{split} \label{eqn:testmatmul_complex}
\end{equation}

Our complex matrix multiplication was implemented using three real matrix multiplications via the 3M method, a variant of the Karatsuba algorithm~\cite{kouya_iccsa2024} as follows:
\begin{equation}
\begin{split}
	C_1 &:= \mbox{Re}(A)\mbox{Re}(B),\ C_2 := \mbox{Im}(A)\mbox{Im}(B) \\
	\mbox{Re}(C) &:= C_1 - C_2 \\
	\mbox{Im}(C) &:= (\mbox{Re}(A)+\mbox{Im}(A))(\mbox{Re}(B)+\mbox{Im}(B)) - C_1 - C_2.
\end{split} \label{eqn:cgemm_3m}
\end{equation}
Therefore, both the Strassen- and block-based algorithms perform three real matrix multiplications. The numerical accuracy, measured as the minimum of the real and imaginary parts, falls within the ranges of 23.2--24.4 significant decimal digits for DD, 39.2--40.8 for TD, and 55.8--56.5 for QD precision.

\begin{figure*}[htb]
\begin{center}
\includegraphics[width=.49\textwidth]{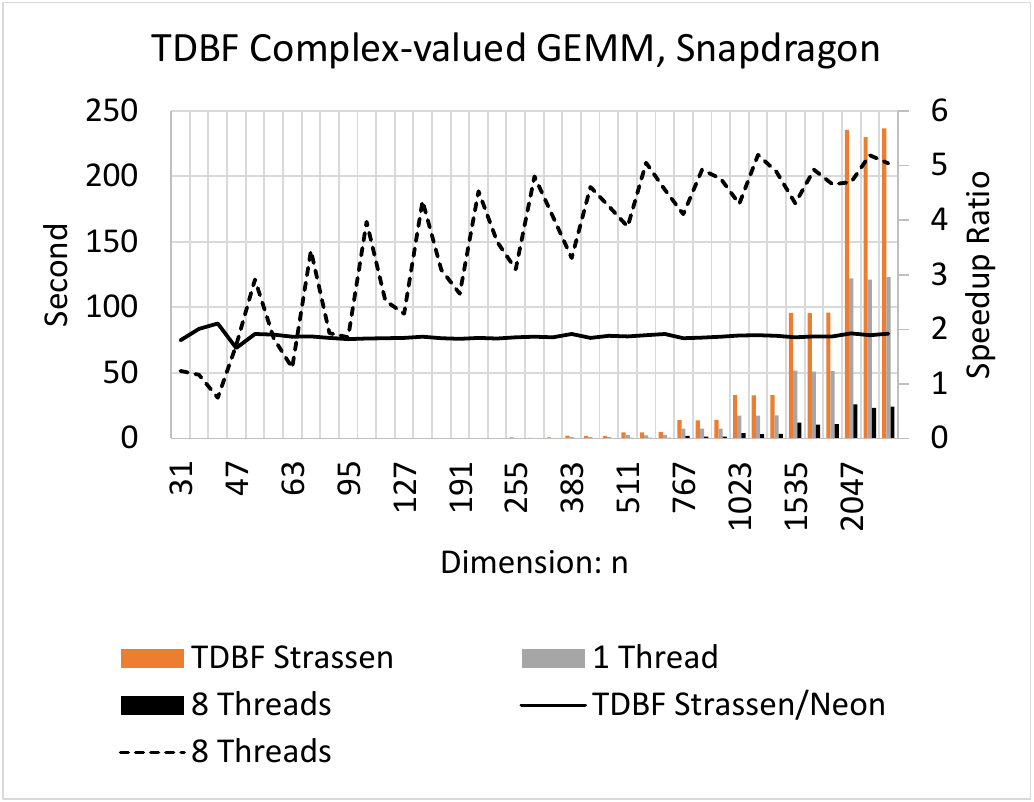}
\includegraphics[width=.49\textwidth]{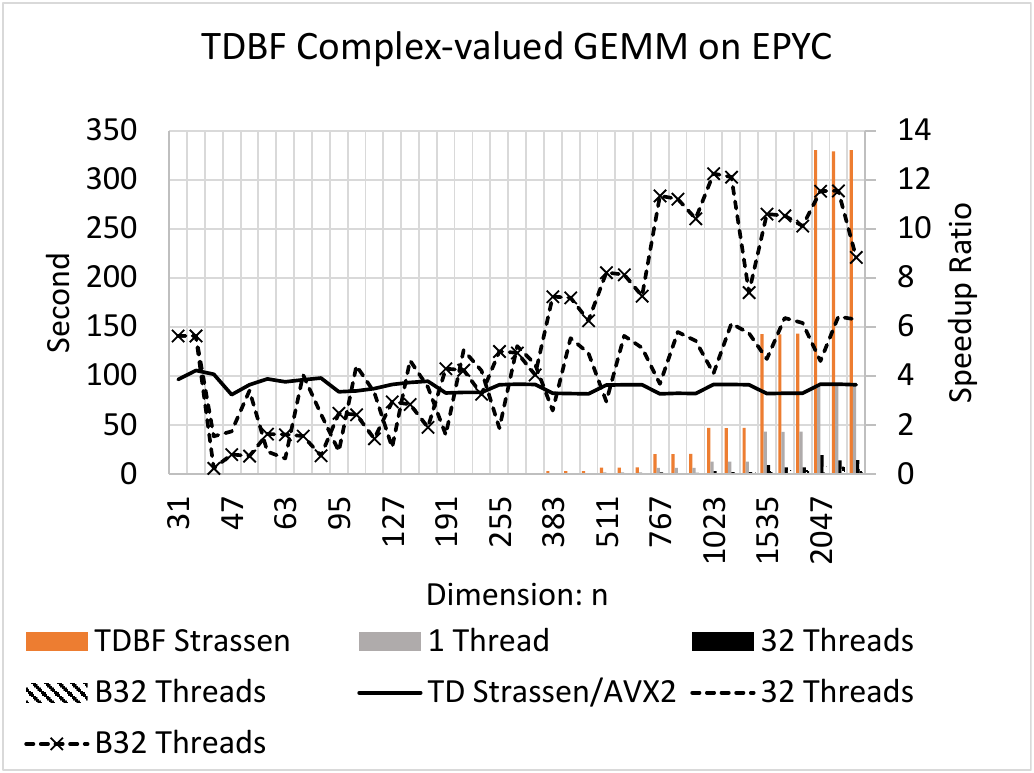}
\includegraphics[width=.49\textwidth]{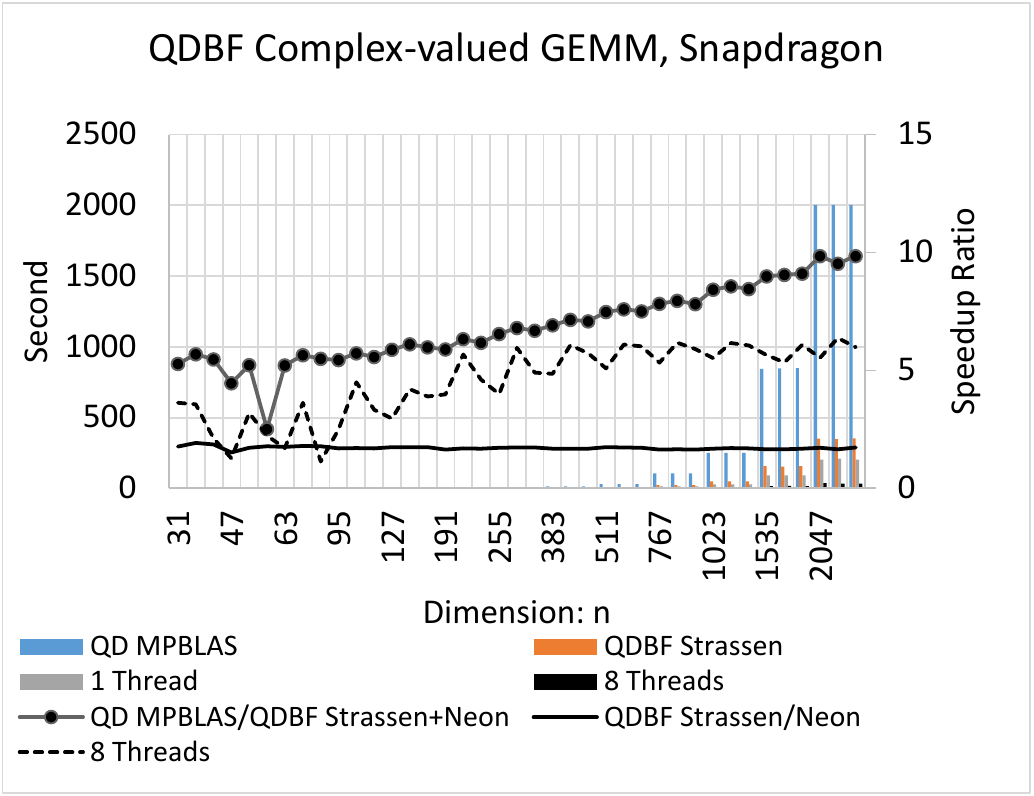}
\includegraphics[width=.46\textwidth]{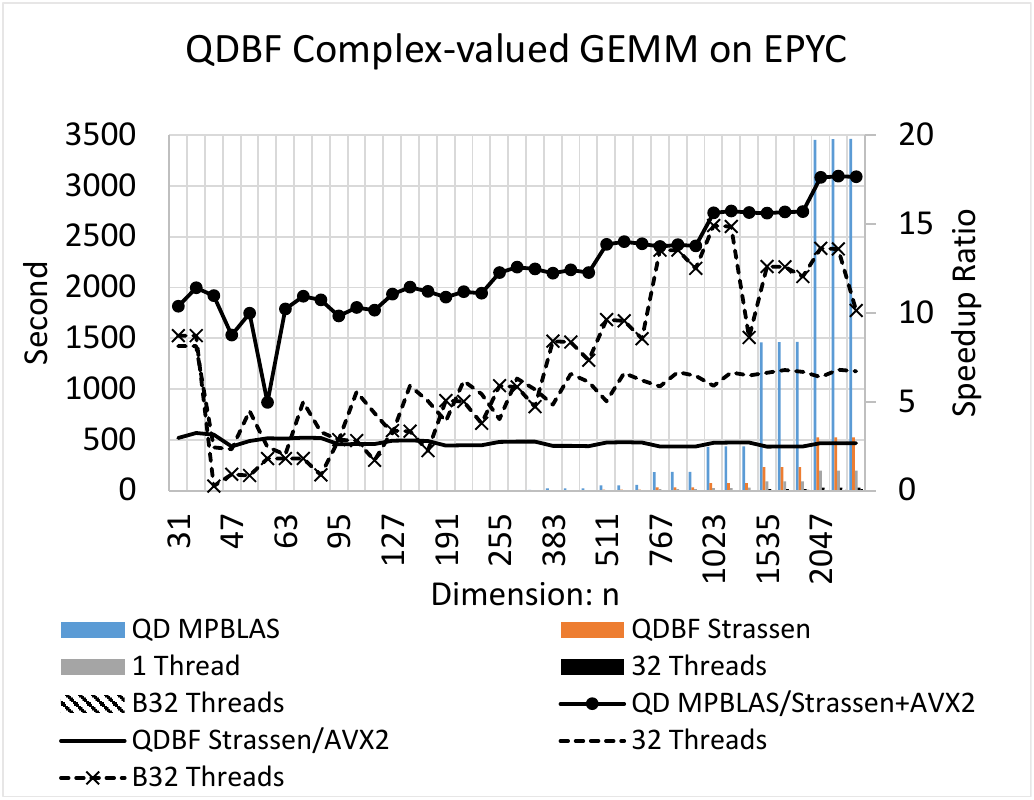}
\caption{Computation time (s) of complex-valued Strassen matrix multiplication: Snapdragon (left), EPYC (right)}\label{fig:cmatmul_strassen_bf}
\end{center}
\end{figure*}

Figure~\ref{fig:cmatmul_strassen_bf} presents the benchmark results for complex matrix multiplication on the EPYC and Snapdragon platforms.

The results demonstrate performance characteristics consistent with those observed for real matrix multiplication. In particular, extracting the computation times for $n = 2049$ reveals that our implementation achieved speedups that significantly exceeded those for real matrix multiplication relative to MPBLAS, as listed in Table~\ref{table:cmblas_speedup}.

\begin{table}[htb]
\centering
		\caption{Complex-valued matrix multiplication time (s) for $n=2049$}\label{table:cmblas_speedup}
\begin{tabular}{ll|r|r|r|r|r}
\toprule
Precision & Platform & MPBLAS & S+SIMD & 8 Threads & 32 Threads & B32 Threads \\
\midrule
\multirow{2}{*}{DD}
 & Snapdragon & 72.1 & 29.9 & 7.8 & -- & -- \\
 & EPYC & 254.3 & 20.4 & 5.0 & 4.4 & 3.0 \\
\midrule
\multirow{2}{*}{TDBF}
 & Snapdragon & -- & 123.2 & 24.4 & -- & -- \\
 & EPYC & -- & 90.5 & 14.6 & 14.4 & 10.3 \\
\midrule
\multirow{2}{*}{QDBF}
 & Snapdragon & 2004.2 & 203.4 & 34.0 & -- & -- \\
 & EPYC & 3463.1 & 196.1 & 29.2 & 29.2 & 19.4 \\
\bottomrule
\end{tabular}
\end{table}

Consequently, our complex matrix multiplication requires only 2.8--3.5$\times$ the computation time of real matrix multiplication across all precision levels. In contrast, MPBLAS, which employs the standard 4M method for each operation, requires 3.5--18.4$\times$ the computation time. These results indicate that the 3M-based implementation becomes increasingly advantageous as precision and matrix size increase.

%
\section{Benchmark Tests for Polynomial Evaluation and Algebraic Equation Solvers}

As demonstrated by the matrix multiplication benchmarks, BF arithmetic offers no speedup in DD precision when conventional sloppy addition and multiplication are employed. In this section, we present the benchmark results for a polynomial evaluation and an algebraic equation solver based on the Durand--Kerner (DK) method, both of which utilize TD and QD precision arithmetic, for which performance improvements are anticipated. As described previously, we did not use the BF algorithm for the DD evaluation of a polynomial, but used the BF polynomial evaluation for TD and QD arithmetic.

%
\subsection{Evaluation of Real-Coefficient Polynomial Functions}

For the polynomial evaluation, the coefficients of the degree-$n$ real-coefficient polynomial $p_n(x) = \sum^n_{i=0} a_i x^i$ were generated randomly, and the results were compared among the standard Horner's method, Estrin's method, and SIMD-vectorized Estrin's method.

\begin{figure}[htb]
\begin{center}
\includegraphics[width=.84\textwidth]{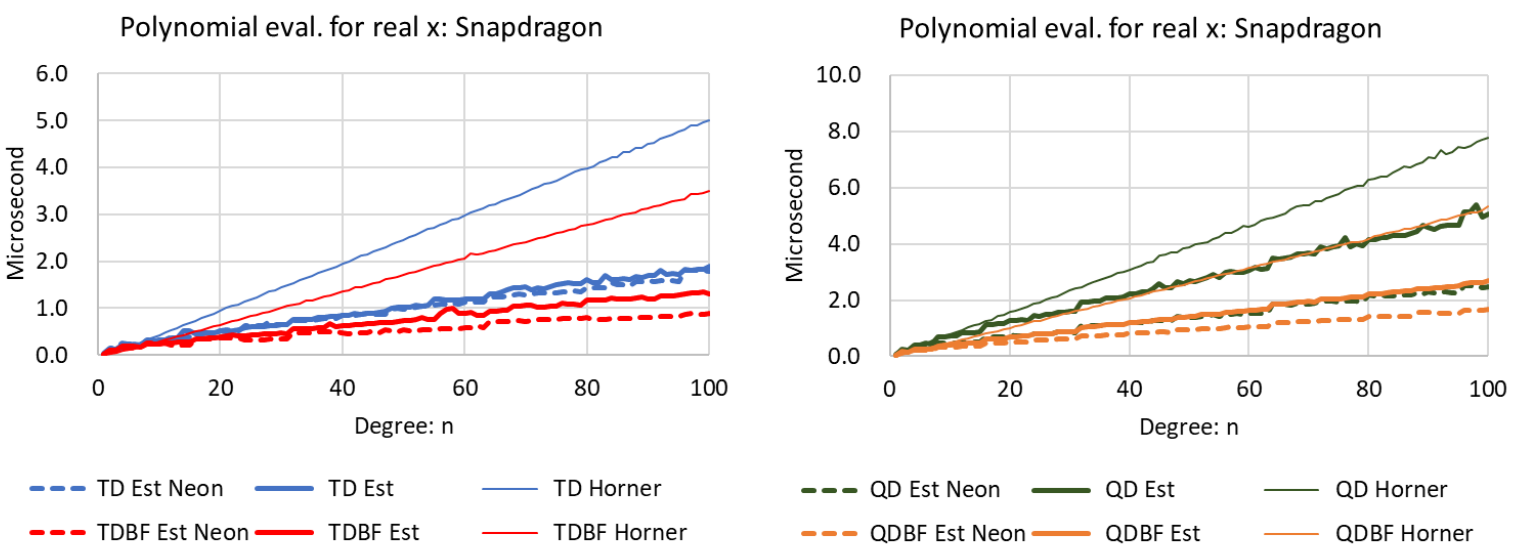}
\includegraphics[width=.84\textwidth]{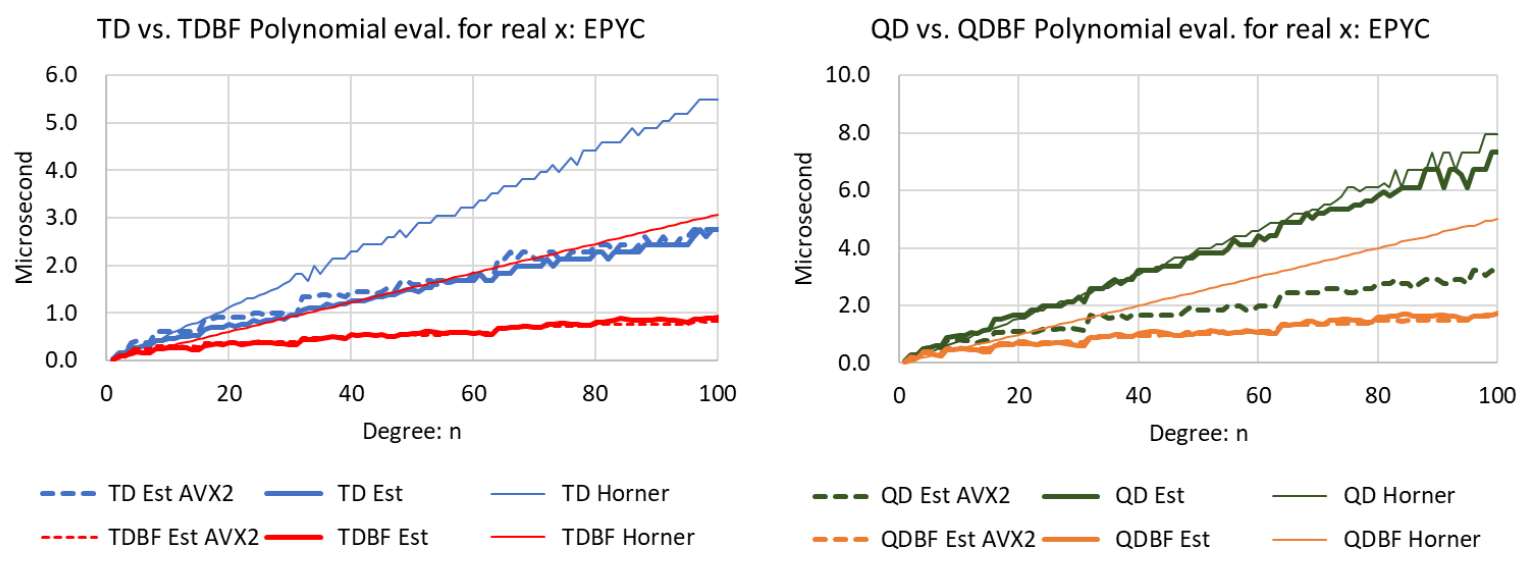}
\caption{Computation time ($\mu$s) for evaluating a real-coefficient polynomial $p_n(x)$ at real arguments: Snapdragon (top), EPYC (bottom)}\label{fig:polyval_r}
\end{center}
\end{figure}

Figure~\ref{fig:polyval_r} shows the computation time (in microseconds) required to evaluate the real-coefficient polynomial $p_n(x)$ in real arguments. The upper panel corresponds to the Snapdragon platform and the lower panel to the EPYC platform; the standard arithmetic (NoBF) is compared with branch-reduction BF arithmetic (BF). As the degree (or equivalently, the evaluation count corresponding to the problem size) increases, the computation time increases, and the effects of branch elimination and vectorization become more pronounced, particularly for high-precision variants such as TD and QD, where the operation count and cost are dominant. Across both architectures, a consistent trend is observed: BF reduces the computation time relative to NoBF for TD and QD precisions.

\begin{figure}[htb]
\begin{center}
\includegraphics[width=.84\textwidth]{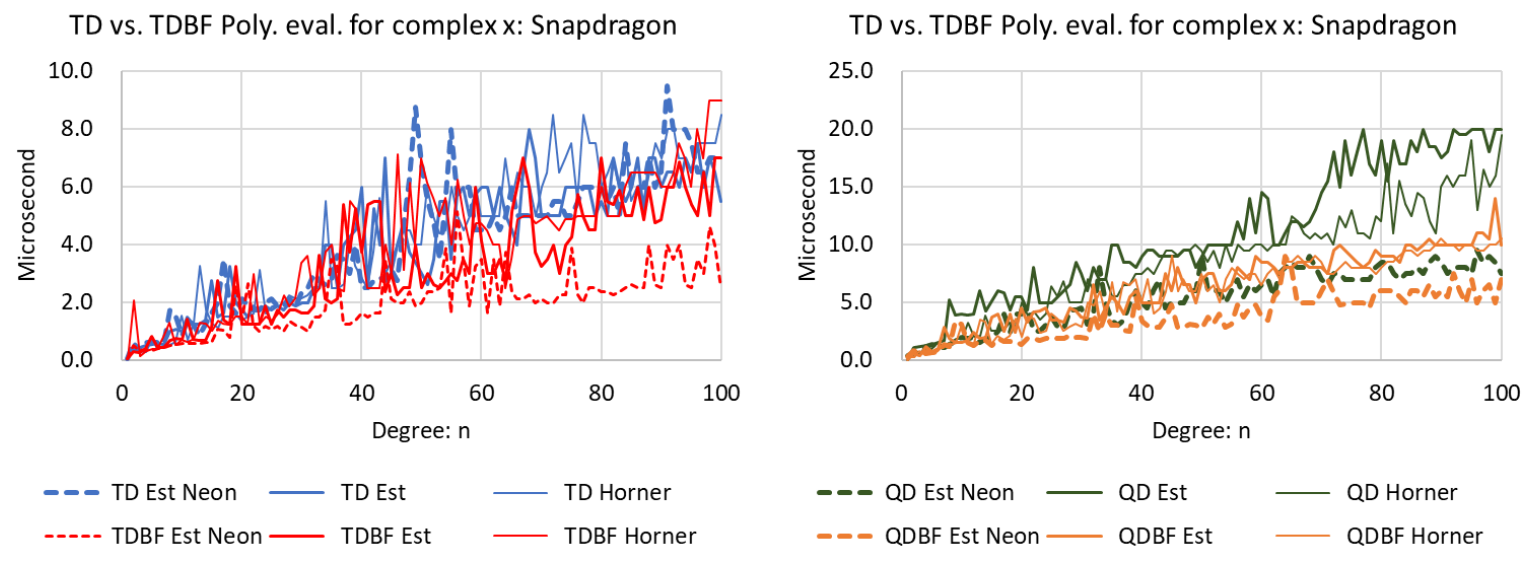}
\includegraphics[width=.84\textwidth]{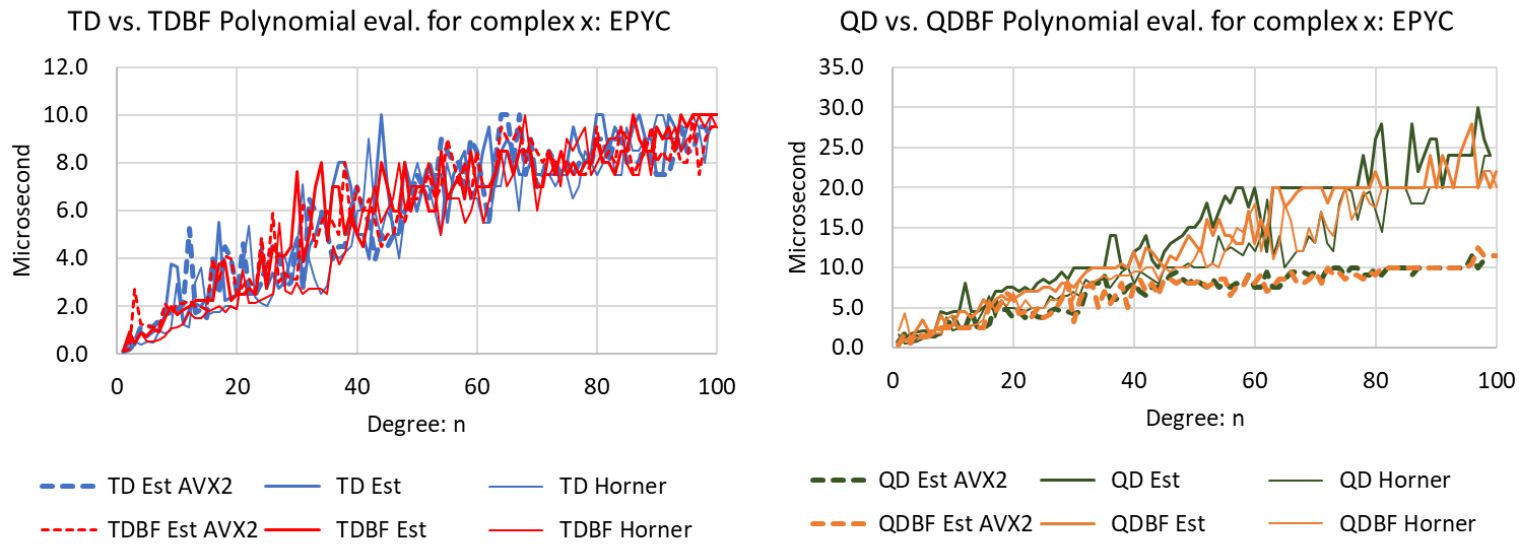}
\caption{Computation time ($\mu$s) for evaluating a real-coefficient polynomial $p_n(x)$ at complex arguments: Snapdragon (top) and EPYC (bottom)}\label{fig:polyval_c}
\end{center}
\end{figure}

Figure~\ref{fig:polyval_c} shows the computation time (in microseconds) required to evaluate the real-coefficient polynomial $p_n(x)$ for the complex arguments. Because each complex multiplication and addition decomposes into multiple real operations, the computation times are higher than those for real-argument evaluation, as shown in Fig. ~\ref{fig:polyval_r}. In contrast, as the number of operations increases, the impact of the conditional branching and normalization steps also increases; therefore, BF arithmetic should, in principle, remain effective for complex-argument evaluation. Although the Snapdragon results largely conform to this expectation, the EPYC results show a slight discernible difference. In particular, the performance difference for TD arithmetic appears to be minimal, even when one exists.

%
\subsection{Algebraic Equation Solver Based on the DK Method}

The method being compared is the quadratically convergent DK method, which is a classical simultaneous iterative solver. The approximate roots after $k$ iterations are expressed as follows:
\[ \mathbf{z}_k = [z^{(k)}_1\ z^{(k)}_2\ ...\ z^{(k)}_n]^T \in \mathbb{C}^n, \]
The recurrence relationship, represented using the monic polynomial $q_n(x) = p_n(x)/a_n$, is expressed by
\begin{equation}
 z^{(k+1)}_i := z^{(k)}_i - \frac{q_n(z^{(k)}_i)}{\prod^n_{j=1, j\not=i} (z^{(k)}_i - z^{(k)}_j)}\label{eqn:dk2}
\end{equation}
as the quadratically convergent DK method.

The Aberth initial values are expressed as
\begin{equation}
	z^{(0)}_i := -\frac{c_{n-1}}{n} + r\exp\left\{\left(\frac{2(i-1)\pi}{n}+\frac{3}{2n}\right)\mathrm{i}\right\} \label{eqn:dka_init}
\end{equation}
In the experiments, the radius $r$ was defined using the number of nonzero coefficients $n_{\rm nz} \le n$ as
\[ r := \max_{0 \le i \le (n-1)} |n_{\rm nz}\, c_i|^{1/(n-i)}. \]
The test problem was the Chebyshev integration problem, in which the polynomial coefficients with $a_n = 1$ were computed as follows:
\begin{equation}
\begin{cases}
	a_{n-(2k-1)} := 0 & \\
	a_{n - 2k} := -\sum^k_{2j+1} a_{n-2(k-j)} & \\
\end{cases}\label{eqn:chebcoef}
\end{equation}
where $k = 1, 2, \ldots, \lfloor n/2 \rfloor$.

Table~\ref{table:cheby64} presents the benchmark results obtained by applying the DK method with DD, TD, and QD precision arithmetic to the Chebyshev integration problem, as described previously.
\begin{table}[htb]
\begin{center}
\caption{Benchmark test (unit: second) for the Chebyshev integration problem ($n=64$)}\label{table:cheby64}
\begin{tabular}{c|c|c|c}\toprule
Snapdragon & DD 		& TD 	& QD \\ \midrule
NoBF 		& 0.0528 & 0.174 & 0.59 \\
NoBF Neon	& 0.0381 & 0.172 & 0.47 \\ \midrule
BF		 	& --- & 0.14 & 0.383 \\
BF Neon		& --- & 0.111 & 0.336 \\ \bottomrule
\end{tabular}
\begin{tabular}{c|c|c|c}\toprule
EPYC		& DD 		& TD 	& QD \\ \midrule
NoBF 		& 0.0627	& 0.306 & 0.989 \\
NoBF AVX2	& 0.0431	& 0.251 & 0.608 \\ \midrule
BF		 	& ---	& 0.214 & 0.628 \\
BF AVX2		& ---	& 0.156 & 0.415 \\ \bottomrule
\end{tabular}
\end{center}
\end{table}

Table~\ref{table:cheby64} lists the benchmark results for the Chebyshev integration problem ($n = 64$). The results with and without SIMD vectorization---Neon on the Snapdragon platform and AVX2 on the EPYC platform---are listed side-by-side (NoBF Neon/AVX2 and BF Neon/AVX2), and the effects of BF conversion (BF) are compared for TD and QD precisions. The effect of BF conversion is pronounced for TD and QD; on the EPYC, BF achieves an approximately 1.43$\times$ speedup for TD and 1.57$\times$ for QD relative to NoBF, and even when combined with SIMD (NoBF AVX2$\rightarrow$BF AVX2), speedups of approximately 1.61$\times$ for TD and 1.47$\times$ for QD are obtained. We can observe similar trends for the Snapdragon, where the combination of BF conversion and Neon vectorization is effective for TD and QD; in particular, TD achieves an approximately 1.55$\times$ speedup in the NoBF Neon$\rightarrow$BF Neon transition. These results confirm that the combined use of BF conversion and SIMD vectorization is effective for multicomponent multiple-precision arithmetic with TD and QD precisions.

%
\section{Conclusion and Future Work}
\label{sec:conclusion}

In this study, we applied BF algorithms that suppress conditional branching to multicomponent multiple-precision arithmetic (DD/TD/QD) and evaluated their performance on CPU SIMD (AVX2 on EPYC) and Arm Neon. Benchmark tests were conducted for real and complex square matrix multiplications, real- and complex-valued evaluations of real-coefficient polynomials, and the Chebyshev integration problem. The results confirm that, for high precision (particularly TD and QD), the combined use of BF conversion and SIMD vectorization is effective and reduces the computation time compared with the standard implementation (NoBF). In addition, the results validate that BF conversion offers no benefit to DD precision arithmetic. These findings indicate that achieving performance improvements in multicomponent arithmetic requires (i) branch elimination (BF conversion) and (ii) the exploitation of instruction-level parallelism (SIMD vectorization) to be combined with an implementation strategy selected according to the target problem, precision level, and architecture.

The following directions are identified for future research:
\begin{enumerate}
 \item \textbf{Adaptive implementation selection (autotuning)}:
 As the optimal execution path (NoBF/BF, SIMD width, and loop partitioning) varies with the precision level (DD/TD/QD) and problem size, we will introduce a mechanism for selecting the optimal implementation at either runtime or build time.

 \item \textbf{Support for wider SIMD extensions and instruction sets}:
 Extensions to AVX-512 and Arm SVE/SVE2 will be pursued, along with establishing design guidelines for data layout and reduction operations as register width increases.

 \item \textbf{Memory access optimization and enhanced parallelization}:
 In addition to improving cache efficiency in matrix operations and polynomial evaluation, we will optimize the coordination of thread-level parallelism via OpenMP with SIMD vectorization, including NUMA placement and scheduling.


 \item \textbf{Extension to application problems}:
 The scope of the application will be expanded to include high-precision linear algebra problems (QR decomposition, eigenvalue problems, and iterative methods), special function evaluation, numerical integration, and optimization, and the effectiveness in practical settings (convergence behavior, total runtime, and energy efficiency) will be evaluated.
\end{enumerate}

Through these efforts, we aim to realize accelerated multicomponent high-precision arithmetic in a more general and portable form by establishing a high-precision numerical computing infrastructure applicable to a wide range of computational environments, ranging from mobile devices to supercomputers.

%
\section*{Acknowledgment}

This research was supported by JSPS KAKENHI (grant number 23K11127).


\begin{thebibliography}{10}

\bibitem{qd}
D.H. Bailey.
\newblock {QD}.
\newblock \url{https://www.davidhbailey.com/dhbsoftware/}.

\bibitem{dekker}
T.~J. Dekker.
\newblock A floating-point technique for extending the available precision.
\newblock {\em Numerische Mathematik}, Vol.~18, No.~3, pp. 224--242, Jun 1971.

\bibitem{mpc}
Andreas Enge, Philippe Th\'eveny, and Paul Zimmermann.
\newblock {MPC}.
\newblock \url{http://www.multiprecision.org/mpc/}.

\bibitem{triple_word_prec2019}
N.~{Fabiano}, J-.M. {Muller} and J.{Picot}.
\newblock Algorithms for triple-word arithmetic.
\newblock {\em IEEE Trans. on Computers}, Vol.~68, pp. 1573--1583, 2019.

\bibitem{gmp}
T.~Granlaud and {GMP} development team.
\newblock The {GNU} {M}ultiple {P}recision arithmetic library.
\newblock \url{https://gmplib.org/}.

\bibitem{dd_avx_original}
Toshiaki Hishinuma, Akihiro Fujii, Teruo Tanaka, and Hidehiko Hasegawa.
\newblock AVX acceleration of DD arithmetic between a sparse matrix and vector.
\newblock In Roman Wyrzykowski, Jack Dongarra, Konrad Karczewski, and Jerzy
 Wa{\'{s}} Niewski, editors, {\em Parallel Processing and Applied Mathematics}
 pp. 622--631, Berlin, Heidelberg, 2014. Springer Berlin Heidelberg.

\bibitem{lis}
T.Kotakemori, S.Fujii, H.Hasegawa, and A.Nishida.
\newblock Lis: Library of iterative solvers for linear systems.
\newblock \url{https://www.ssisc.org/lis/}.

\bibitem{bncmatmul}
Tomonori Kouya.
\newblock {BNC}matmul.
\newblock \url{https://github.com/tkouya/bncmatmul}.

\bibitem{kouya_iccsa2021}
Tomonori Kouya.
\newblock Acceleration of multiple precision matrix multiplication based on
 multicomponent floating-point arithmetic using avx2.
\newblock In Osvaldo Gervasi, Beniamino Murgante, Sanjay Misra, Chiara Garau,
 Ivan Ble{\v{c}}i{\'{c}}, David Taniar, Bernady~O. Apduhan, Ana Maria A.~C.
 Rocha, Eufemia Tarantino, and Carmelo~Maria Torre, editors, {\em
 Computational Science and Its Applications -- ICCSA 2021}, pp. 202--217,
 Cham, 2021. Springer International Publishing.

\bibitem{kouya_iccsa2024}
Tomonori Kouya.
\newblock Performance evaluation of^^c2^^a0accelerated complex
 multiple-precision lu decomposition.
\newblock In Osvaldo Gervasi, Beniamino Murgante, Chiara Garau, David Taniar,
 Ana Maria~A. C.~Rocha and Maria~Noelia Faginas~Lago, editors, {\em
 Computational Science and Its Applications, ICCSA 2024 Workshops}, pp.
 3--19, Cham, 2024. Springer Nature, Switzerland.

\bibitem{pair_arithmetic}
Marko Lange and Siegfried~M. Rump.
\newblock Faithfully rounded floating-point computations.
\newblock {\em ACM Trans. Math. Softw.}, Vol.~46, No.~3, July 2020.

\bibitem{GQD}
Mian Lu, Bingsheng He, and Qiong Luo.
\newblock Supporting extended precision on graphics processors.
\newblock In {\em Proceedings of the Sixth International Workshop on Data
 Management on New Hardware}, DaMoN '10, pp. 19--26, New York, NY, USA, 2010.
 ACM.

\bibitem{mplapack}
MPLAPACK/MPBLAS.
\newblock Multiple precision arithmetic {LAPACK} and {BLAS}.
\newblock \url{https://github.com/nakatamaho/mplapack}.

\bibitem{pair_arith_sparse2025}
Daichi Mukunoki and Katsuhisa Ozaki.
\newblock Sparse iterative solvers using high-precision arithmetic with quasi
 multi-word algorithms.
\newblock In {\em 2025 IEEE 18th International Symposium on Embedded
 Multicore/Many-core Systems-on-Chip (MCSoC)}, pp. 33--40, 2025.

\bibitem{mpfr}
{MPFR} Project.
\newblock The {MPFR} library.
\newblock \url{https://www.mpfr.org/}.

\bibitem{automatic_verification_2025}
David~K. Zhang and Alex Aiken.
\newblock Automatic verification of^^c2^^a0floating-point accumulation
 networks.
\newblock In Ruzica Piskac and Zvonimir Rakamari{\'{c}}, editors, {\em Computer
 Aided Verification}, pp. 215--237, Cham 2025. Springer Nature, Switzerland.

\bibitem{branch_free_2025}
David~Kai Zhang and Alex Aiken.
\newblock High-performance branch-free algorithms for extended-precision
 floating-point arithmetic.
\newblock In {\em Proceedings of the International Conference for High
 Performance Computing, Networking, Storage and Analysis}, SC '25, p.
 695^^e2^^80^^93710, New York, NY, USA, 2025. Association for Computing
 Machinery.

\bibitem{kouya_arm_neon_2025}
Tomonori Kouya.
\newblock Trial approach to accelerate multi-component-type multiple-precision basic linear computation with Arm neon intrinsics (in Japanese).
\newblock Technical report, HPC,
 Sep 2025.

\end{thebibliography}

\end{document}